\documentclass[journal,comsoc]{IEEEtran}
\usepackage[T1]{fontenc}

\usepackage{booktabs}
\usepackage{graphicx}
\usepackage{array}
\usepackage{float}
\usepackage{multirow}
\usepackage{subfigure} 
\usepackage{CJK}
\usepackage{color}

\ifCLASSINFOpdf
\else
\fi
\usepackage{amsmath}
\usepackage[cmintegrals]{newtxmath}
\begin{document}

%
\title{Agent-Based Campus Novel Coronavirus Infection and Control Simulation}
%
%
%

\author{Pei~Lv,
	Quan~Zhang, Boya~Xu, Ran~Feng, ChaoChao~Li, 
	Junxiao Xue, 
	Bing~Zhou,
        and~Mingliang~Xu,~\IEEEmembership{Member,~IEEE}
\thanks{Pei Lv, Quan Zhang, Boya Xu, Ran Feng, Chaochao Li, Bing Zhou and Mingliang Xu are with the School of Information Engineering, Zhengzhou University, Zhengzhou 450001, China (E-mail: ielvpei@zzu.edu.cn; quanzhang@gs.zzu.edu.cn; xuboya@gs.zzu.edu.cn; fengrancele@163.com; ieccli@zzu.edu.cn; iebzhou@zzu.edu.cn; iexumingliang@zzu.edu.cn). Junxiao Xue is with the School of Software, Zhengzhou University, Zhengzhou 450001, China (E-mail: xuejx@zzu.edu.cn.)}
}

\maketitle

\begin{abstract}
Corona Virus Disease 2019 (COVID-19), due to its extremely high infectivity, has been spreading rapidly around the world and bringing huge influence to socioeconomic development as well as people's daily life. Taking for example the virus transmission that may occur after college students return to school, we analyze the quantitative influence of the key factors on the virus spread, including crowd density and self-protection. One Campus Virus Infection and Control Simulation model (CVICS) of the novel coronavirus is proposed in this paper, fully considering the characteristics of repeated contact and strong mobility of crowd in the closed environment. Specifically, we build an agent-based infection model, introduce the mean field theory to calculate the probability of virus transmission, and micro-simulate the daily prevalence of infection among individuals. The experimental results show that the proposed model in this paper efficiently simulate how the virus spread in the dense crowd in frequent contact under closed environment. Furthermore, preventive and control measures such as self-protection, crowd decentralization and isolation during the epidemic can effectively delay the arrival of infection peak and reduce the prevalence, and 
finally lower the risk of COVID-19 transmission after the students return to school.
\end{abstract}


%
\IEEEpeerreviewmaketitle

\section{Introduction}
\begin{figure}
	\centering
	\subfigure[Keeping distances]{
		\begin{minipage}[t]{0.49\columnwidth}
			\centering
			\includegraphics[width=1.7in,height=1in]{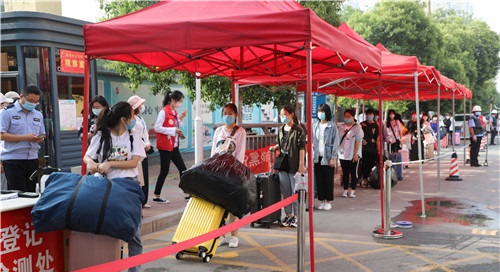}
		\end{minipage}%
	}%
	\subfigure[Taking temperature]{
		\begin{minipage}[t]{0.49\columnwidth}
			\centering
			\includegraphics[width=1.7in,height=1in]{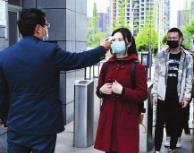}
		\end{minipage}
	}%
	\centering
	\caption{Some of campus prevention and control measures.}
	\label{fig1}
\end{figure}

%
%
%
%
\IEEEPARstart{A}{t} the end of 2019, the pneumonia epidemic caused by novel coronavirus swept the world, bringing dramatic effects on the economy, production, life, etc \cite{world2020global}. The virus spreads rapidly and widely, especially in the environment with high crowd density and strong crowd mobility. In response to the sudden outbreak, the Chinese Government  took various counter-measures immediately, such as extending citizen holidays, postponing the return to work and school, and assisting the disaster areas, which effectively reduce the impact of the spread of the virus. To tackle the problem of suspending classes, education institutions around the world proposed various distance learning programs through modern technologies to guarantee students to continue their education. However, due to the unsatisfactory online teaching quality, especially for some professional experiments, it is still very important and essential for students to return back to school. As a relatively special group, college students spread all over the country, and will contact with a large number of people on the way back to school. After returning to school, students will inevitably gather together to study and live. The above phenomenons will have great negative impact on the epidemic prevention and control, and the large gatherings of dense crowd with strong mobility will even worsen the epidemic \cite{milne2008small}. Therefore, how to manage the college students after they return back to school is one of the most significant problems to be solved. Figure~\ref{fig1} shows several common prevention and control measures adopted when students enter the campus.

\par Regarding the spread of novel coronavirus, we can analyze their laws and trends by constructing infection models. In most of existing methods, SIR model and SEIR model, which are based on the dynamic systems \cite{li1995global}, are often adopted to fit the urban infection curve. However, it is one obviously different situation on campus.  Universities are relatively closed spaces, where plenty of contact is unavoidable  \cite{ridenhour2011controlling}. The average number of people contacted each day on campus is larger than that of urban residents. The existing urban simulation model cannot be applied to the university environment directly, since the novel coronavirus is quite different from other previous infectious diseases. For example, SARS and another infamous coronavirus, which broke out in 2003, researchers have also built corresponding models to study its infectious characteristics. Although the incidence characteristics and other aspects of SARS are similar to COVID-19, SARS does not have infectiousness in its latent period, which is very different from the present one. Moreover, the existing simulation models for large scale population are usually macroscopic, which means they do not focus on individuals, hence cannot reflect the differences among individuals. The microscopic simulation model requires the consideration of attributes of each individual in detail. As the number of individuals increases, the calculation time will increases significantly. Considering the virus spreads rapidly and widely, the computational efficiency of the model is as important as its accuracy.
\par In response to the outbreak of coronavirus, governments around the world have taken various measures to lower the influence. As an example, China has developed a series of containment and  obtained satisfying results. To be specific, the Chinese government approved the travel ban, encouraged people to stay at home to avoid gathering, restricted social activities, etc. However, some countries or organizations ignore the spread effect of the virus in the dense crowd, which aggravates the situation. A case in point is "Diamond Princess", where the virus spread rapidly due to insufficient preventive and control measures in the early stage. Taking a lesson from it, it is necessary to take strict measures to curb the spread of the virus in the environment with high-density population \cite{tabari2020nations}.
\par  In this paper, we comprehensively consider various factors influencing the spread of the disease, and propose a campus virus infection and control simulation (CVICS) model of the novel coronavirus, considering the characteristics of repeated contact and strong mobility of the crowd in the closed environment. The agent is the primary component in simulation, and each individual can perceive the surrounding environment independently \cite{fischer2009gpu}. The movements of the individual are driven by the social force \cite{1995Social}\cite{zanlungo2011social}. At the same time, the spread of the virus among individuals in different scenarios on campus will be simulated. The advantage of the proposed model is that it can present the state information of each agent in the environment at any moment from the micro perspective, making the simulation show more details. Taking the crucial factors such as crowd density and self-protection into account, effective preventive and control measures such as travelling in batches, staggered travel and isolation are put forward. According to several groups of comparative experiments, we can observe the trend of virus transmission on campus under different conditions. Through the intervention in group behavior, the infection rate can be effectively reduced.
\par Our major contributions are listed as follows:
\begin{itemize}
    \item We propose a CVICS model of the novel coronavirus, based on the characteristics of repeated contact and strong mobility of the crowd in the closed environment. Based on this model, we simulate the daily  prevalence of infection among college students, and propose effective control measures with qualitative and quantitative analysis. 
    \item Considering the differences among individuals, we design an agent-based simulation method, introduce the mean field theory to calculate the probability of virus transmission, and micro-simulate the daily prevalence of infection among individuals efficiently.
    \item Taking Zhengzhou University as an example, we simulate the virus infection among college students and staff during the epidemic period. And we put forward some control measures such as travelling in batches, staggered shifts and isolation, and validate their effectiveness on reducing the virus infection rate.
\end{itemize}


\section{Related work}
Our work for novel coronavirus infection simulation is closely related to virus infection model and crowd simulation. Therefore, we mainly discuss some representative work in these two research areas.
\subsection{Infection model}
Nowadays, the spread of virus is generally studied by constructing various infection models, which aims to carry out an accurate risk assessment. In the micro perspective, the classic SIR model divides the population into susceptible, infected, and recovering groups \cite{towers2009pandemic}. It is a good demonstration of the process of infectious diseases from the onset to the end. The premise of using this model is that patients suffering from such infectious diseases can recover, and produce permanent antibodies, and no longer participate in the spread of the disease. As the types of infectious diseases become more complex, the infectious model is being improved constantly. If the infectious diseases have an incubation period, the improved SEIR model based on the SIR model will be used. The SEIR model divides the population into susceptible, latent, infected and rehabilitated, which can describe the transmission process of the epidemic more accurately. When the SARS epidemic broke out in 2003, most of the SIR or SEIR models were used to fit the infection curve of the people in the city. Liang et al.  \cite{liang2020mathematical} established a propagation growth model by considering the growth rate and the inhibition constant, which showed that the infection rate and its changes over time are the most important factors affecting the spread of SARS. In addition, there are other similar models such as SI \cite{cunde2002si} and SIRS \cite{pang2007delayed}. Shi et al. \cite{yu2018dynamic} proposed an epidemic dynamics model, which can describe the development of the widespread disease situation through numerical simulation. In the study of infectious diseases, researchers not only predict the number of people in various stages of diseases, but also need to consider the causes of diseases, transmission media, other relevant social factors and so on \cite{fisman2014early}. These can help decision-makers formulate prevention and emergency plans to prevent the further deterioration of virus spread. 
\par When dealing with the infection situation of colleges or universities, the recovery state will be not considered. Once the suspected cases are found, they will be sent to the school hospital for isolation immediately and then transferred out of the college to a designated location for treatment. Different from the most commonly used models such as SIR and SEIR, which calculate the trend of the number of people in different states under the epidemic, campus infection simulation focuses on suspected groups that may be infected and tries to prevent the large spread of the virus. Therefore, the most commonly used models at present are not completely suitable for the study of epidemics in universities.

\begin{figure*}[!t] 
	\center{\includegraphics[width=2\columnwidth]  {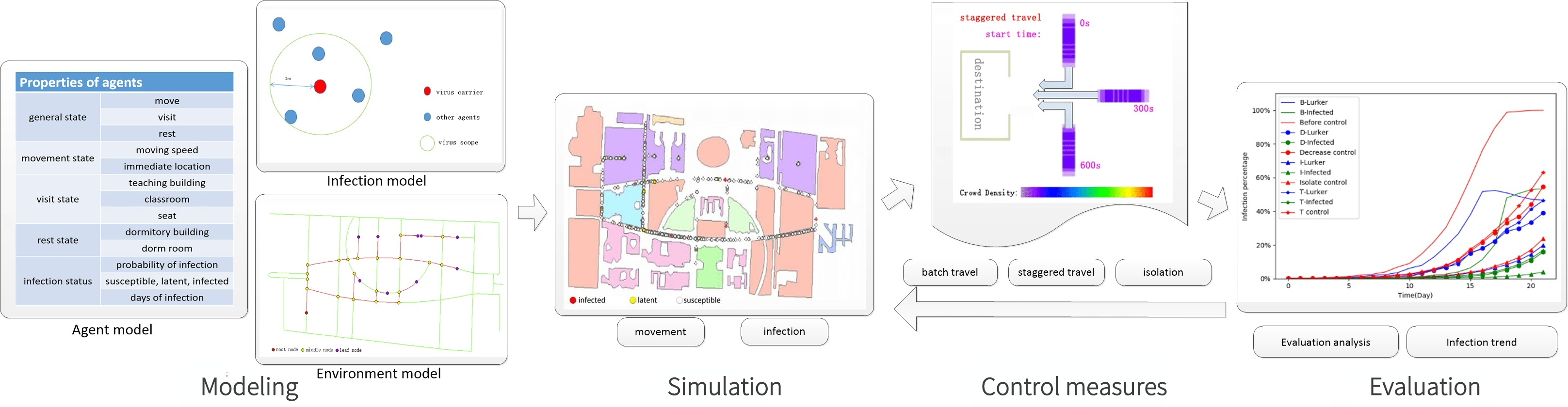}} 
	\caption{\label{fig2} The system overview of our work. Agent model, infection  model and environment model are established for crowd simulation in campus. In the process of simulation, the virus spreads among the moving agent is active in the environment. Furthermore, various control measures are taken according to the simulation results.} 
\end{figure*}

\subsection{Crowd simulation}
The technology of crowd simulation has been widely used in many fields. It can not only be used to generate the simulation animation of the group state \cite{huerre2010simulating}, but also be regarded as an important way to help to design architecture and evaluate the cost~\cite{aschwanden2011empiric}\cite{feng2016crowd}. Common crowd simulation methods include rule-based models, cellular automata models, social force based models \cite{helbing1995social} and agent-based models \cite{huang2009microscopic}. Among them, agent-based simulation has attracted a large amount of attention of many researchers due to its self-adaptive interaction with the environment in complex scenes. In recent years, more and more researchers try to build various simulation models of infectious disease. Kleczkowski et al. \cite{kleczkowski1999mean} used the cellular automata model to explore the impact of relevant information in the local space on the spread of epidemics. Eubank et al. \cite{eubank2004modelling} developed an agent-based epidemic simulation system called  EpiSimdemics. Barrett et al. \cite{barrett2008episimdemics} further developed EpiSimdemics and they proposed a scalable parallel algorithm to simulate the spread of infection in a large-scale social network in reality. Bissett et al. \cite{bissett2016integrated} proposed an integrated method for computational health informatics, which consists of individual-based population construction and agent-based dynamic modeling. At present, there is still a great challenge for large-scale crowd simulation. The macro crowd simulation model usually studies the whole trend and ignores the heterogeneity among individuals. The micro crowd simulation model focuses on single individual, but when the amount increases greatly, the computation will also increase dramatically, and it is difficult to guarantee the speed and accuracy of the simulation. Yang et al. \cite{yang2018mean} recommended the mean-field theory to calculate the crowd movement, which formulates the multi-dimensional problem into two-dimensional to reduce the  computational complexity. It can better handle the simulation problem of large-scale crowd movement. In addition to the human-oriented model mentioned above, vehicles are also  factored into the simulation model. Chao et al.~\cite{2019Force} proposed a novel, extensible, and microscopic method to build heterogeneous traffic simulation in the force-based way. Han et al. \cite{0A} presented a simplified force-based heterogeneous traffic simulation model to facilitate consistent adjustment of the involved parameters. These methods can help to model the environment of virus spread better. 

\par Our model uses an agent-based simulation method to describe microscopic individuals. The agent-based simulation model can vividly describe the impact of the surrounding environment (such as other individuals and obstacles) on the pedestrian behavior, realistically simulate the daily behavior of students on campus, which is not available in macroscopic simulation systems such as EpiSimdemics. By observing the impact of interactions on people in different scenarios (classrooms, restaurants, dormitories, etc.), corresponding measures can be taken to reduce the spread of the virus.

\section{Campus virus infection and control simulation}
The purpose of our model is to study the infection situation in the colleges or universities to assist campus management during the pandemic. We take Zhengzhou University as one typical example. This university is located in the Central Plains of China, where students are widely distributed and the campus environment is very representative.
\par The agent-based model is used to simulate the population infection situation on Zhengzhou University campus during the epidemic under different control measures. In this way, we can ensure the teaching quality while preventing the spread of the virus, and reduce the number of infected people on campus as far as possible. The system overview is shown in Figure 2.

\subsection{Agent definition}
Students in different majors have different learning tasks, so they are divided into four categories according to their majors. In the epidemic environment, each student needs to travel in accordance with the regulations to avoid cross-infection among different groups as shown in Table 1. Each one starts from the dormitory, visits their respective buildings, and returns to the dormitory. According to the Dijkstra \cite{deng2012fuzzy} algorithm, the shortest route between two destinations is obtained. In the simulation, the students walk along the established path. As shown in Figure 3, we use the novel YOLO V5 object detection and DeepSORT object tracking algorithm to track the pedestrians, and then use the  estimated trajectory to calculate the speed of their movement and the Hausdorff distance between them. The walking speed of people in the natural state is approximately normal distribution in the range of 0.926m/s to 1.586m/s. On the premise of large enough space, the distance between two strangers is often kept at 1.55m or more.

\begin{figure}
	\centering
	\subfigure{
		\begin{minipage}[t]{0.49\columnwidth}
			\centering
			\includegraphics[width=1.7in,height=1in]{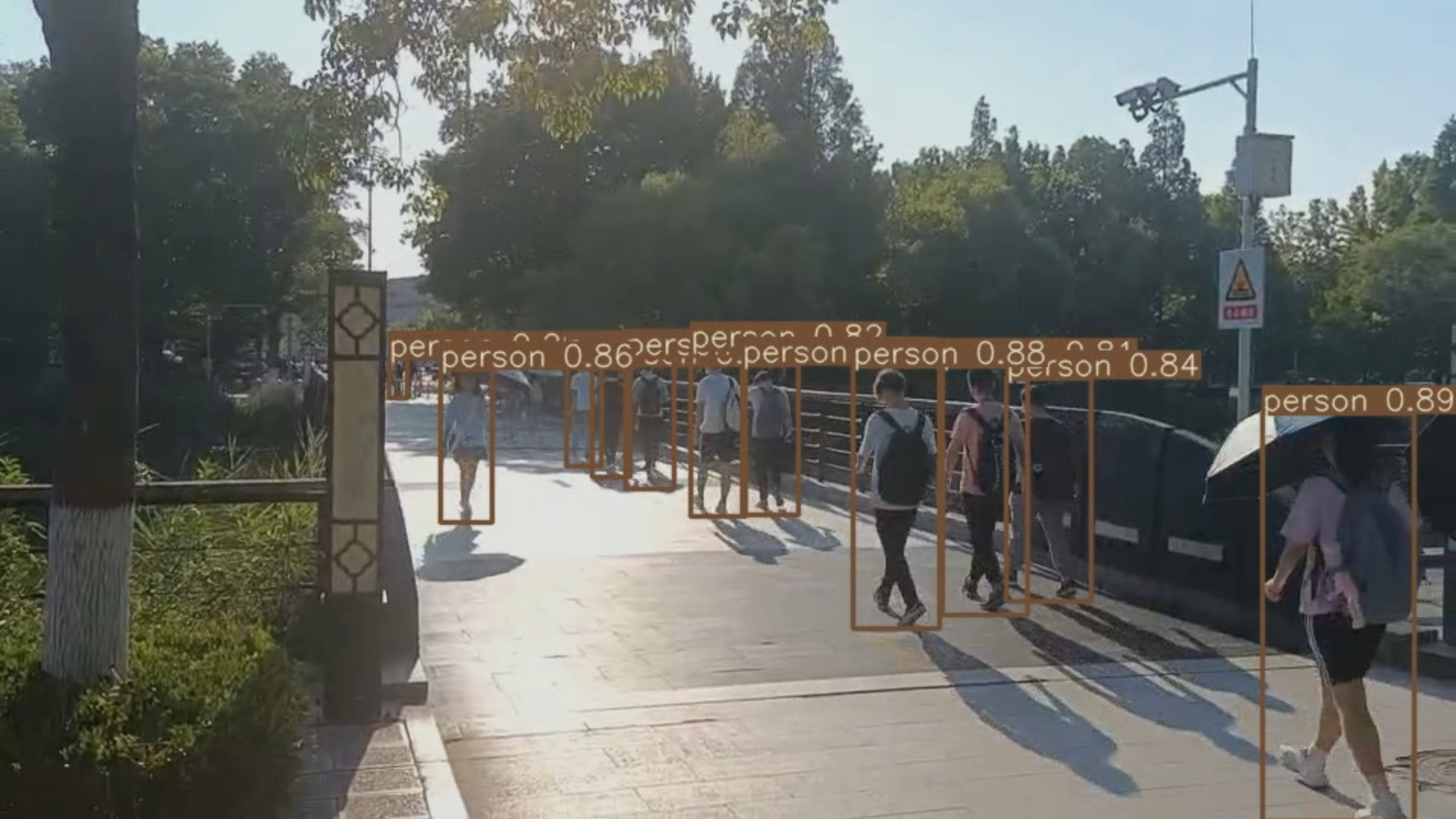}
		\end{minipage}%
	}%
	\subfigure{
		\begin{minipage}[t]{0.49\columnwidth}
			\centering
			\includegraphics[width=1.7in,height=1in]{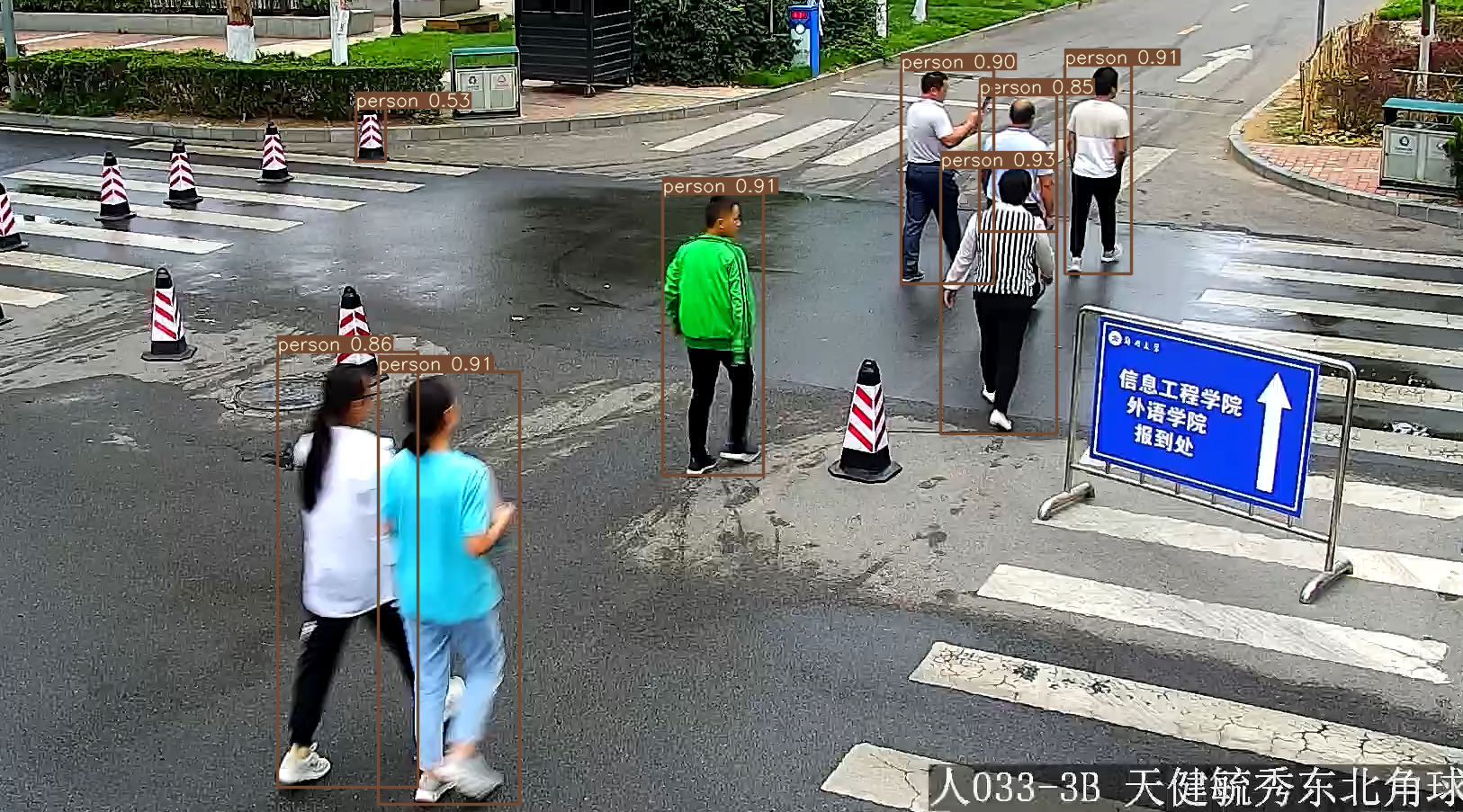}
		\end{minipage}
	}%
	\centering
	\caption{Pedestrian detection from the captured videos on campus.}
\end{figure}

\begin{table}[!t]
	\caption{Walking paths}
	\centering
	\begin{tabular}{c p{6.5cm}}
		\toprule  &Walking Paths\\
		\midrule 
		Category 1 & dormitory$\rightarrow$ teaching \ building\, /\, library $\rightarrow$ restaurant $\rightarrow$ teaching \  building \, /\, library $\rightarrow$ dormitory\\
		Category 2 & dormitory$\rightarrow$ teaching \ building \, /\, laboratory $\rightarrow$ restaurant $\rightarrow$ teaching \ building \, /\, laboratory $\rightarrow$ dormitory\\
		Category 3 &  dormitory$\rightarrow$ laboratory$\rightarrow$ restaurant $\rightarrow$ laboratory $\rightarrow$ dormitory \\
		Category 4 &  dormitory$\rightarrow$ administration\ building \, /\, library $\rightarrow$ restaurant $\rightarrow$ administration\ building \, /\, library $\rightarrow$ dormitory\\
		\bottomrule
	\end{tabular}
\end{table}

\par According to the infection status of one individual, students are divided into three categories: susceptible, latent, and infected. We set a metric $P\_{\mathit{\inf}}$ for the agent to calculate the probability that it may be infected, which can determine the infection status. Within a certain range, if the metric reaches the threshold $T$ ($T=1$), it means that the individual is in an infected state, which changes from susceptible or latent to the infected state. After 7 days, people in the incubation period will be identified as diagnosed patients. According to the transmission characteristics of the novel coronavirus, all students in school are likely to be infected with the virus, and individuals are infectious during the incubation period and the onset period. According to the latest research, the basic reproduction number R0 is as high as 5.7. This means that it is extremely disseminated and infectious when individuals are in latent or infected state~\cite{li2020early}. 
\par In order to simulate the transmission of the virus more realistically, the general state, movement state, visit state, infection state and other attributes are added to each agent, as shown in Table 2.  

\begin{table}[!t]
	\centering
	\caption{Properties of agents}
	\begin{tabular}	{m{3cm}<{\raggedright} m{4.5cm}}
		\hline  Properties&Description\\[3pt]
		\hline \doublerulesep=0.8pt
		general information &gender \\ &age \\
		{general state} & move \\ ~ & visit \\ ~ & rest \\
		{movement state} & moving speed \\ ~ & immediate location \\
		{visit state} &teaching building \\ ~ & classroom \\ ~ & seat \\
		{rest state} & dormitory building \\ & dorm room \\ 
		{infection state} & probability of infection \\ ~ & days of infection \\~ & susceptible, latent, infected \\
		\hline
	\end{tabular}
\end{table}

\subsection{Agent-based novel coronavirus infection modeling}
The infection distance is one of the most important factors in constructing the population infection model. According to \cite{world2020management} and our previous estimated result, the virus carrier can affect other individuals  less than $2m$ away from it.  Meanwhile, the  number of days the virus is carried in the human body also have a vital  impact on the spread of the virus. It is reported that the incubation period of COVID-19 is generally 3 to 22 days, in our experiment, which is set 7 days as that in \cite{fang2020many}. During the incubation period,  virus are also contagious. As the number of days when the human body carry the virus increases, the infectivity increases linearly. The influence of the number of days when a carrier carries the virus on the infectivity of the virus is calculated by the formula (1):
\begin{equation} 
f\left( i_{day} \right)\left. = {\mathit{\min}(}i_{day}/I_{per},1 \right)
\end{equation}
where $i_{day}$ indicates the days when the human carry the virus; $I_{per} = 7$ means that the incubation period is 7 days, and the infectious performance increases linearly along with the increase of the number of days the virus incubates in the human body. When the virus is carried for 7 days or more, the infectivity is no longer enhanced.
\par We consider the influence of the distance between the individual and the  person in the incubation period. The influence of the physical distance is expressed by formula (2):
\begin{equation} 
f\left( d_{n} \right) = \left\{ \begin{array}{l}
{\frac{1}{R}\begin{array}{ll}
	\sqrt{R^{2} - d^{2}} & \\
	\end{array}0 < d < = R} \\
{0\begin{array}{ll}
	& \\
	\end{array}d > R} \\
\end{array} \right.
\end{equation}
where $R$ represents the radius of infection and the value is set to $2m$. When the distance between two individuals $d < = R$, the lager the distance, the smaller the probability of virus transmission. When the distance $d$ among individuals exceeds $R$, the individual will not be infected by the virus.
\par Air humidity, temperature, inhaled air concentration, the mutual distance between individuals and other factors may also have different impacts  on whether a susceptible person will be infected with the virus \cite{liao2005probabilistic}. The combination of these factors from other individuals within the individual's perception range will bring an infection on the central individual. 

\par Therefore, we use the mean field theory to calculate the probability that an individual may be infected with the virus. It comprehensively calculates the impact of all individuals, which can greatly simplify the complexity of the calculation. Specifically, we divide the number of days into 8 time periods when patients carry the virus. They respectively indicate that carrying no virus, carrying the virus for 1 day, and carrying the virus for 7 days or more. The infection probability of one individual is calculated as follows:\\
\begin{equation} 
P\_{\mathit{\inf} =}(\text{1-}\beta){\sum_{i = 0}^{7}{\frac{1}{N}{\sum_{n = 1}^{N}{T_{n}^{j}f\left( i_{day} \right)}}}}f\left( d_{n} \right)
\end{equation}
where $N$ means that there are other $N$ individuals in the perception range of central body;  $T_{n}^{j}$ indicates that the $n$th person carried the virus for $j$ days. The probability distribution is obtained by dividing the number of people in each period by the total number; $\frac{1}{N}{\sum_{n = 1}^{N}T_{n}^{j}}$ indicates the proportion of people in each period; $d_n$ represents the distance between two individuals. $\beta$ represents the group protection rate. In other words, $\beta$ is the proportion of the number of people who take self-protection in all groups.
\par One day is divided into different time slices, and the infection probability is continuously updated with the time slices. During the simulation process, the system calculates the infection rate of the first time slice, then the infection rate of the second time slice, and selects the maximum. At the end of the day, the random value will be assigned automatically. If the value is within the interval of infection probability covered by the previous maximum, it means that the agent will be infected.

\subsection{Simulation environment modeling}

Generally speaking, colleges and universities usually occupy a large area and the roads on the campus are very intricate. To manage the daily activities of teachers and students during the epidemic, they will be prohibited from entering and leaving the campus at will. The campus will be divided into different functional areas, and each area will have only one entrance. The crowd is just allowed to enter and exit from this port, to prevent the spread of viruses caused by the random shuttle of teachers and students in the campus. The buildings in the same functional area are combined as a whole. The winding roads are closed and not passable, as shown in Figure 4.
\begin{figure}[!t] 
	\center{\includegraphics[width=0.9\columnwidth]  {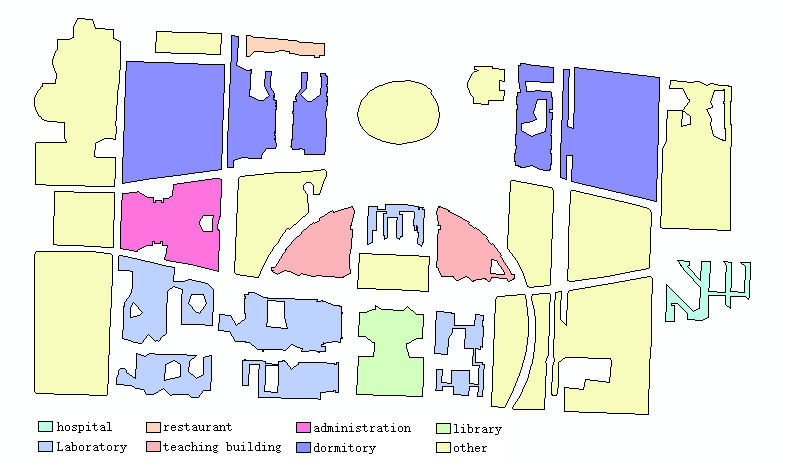}} 
	\caption{\label{1} Building distribution map.} 
\end{figure}
\begin{figure}[!t] 
	\center{\includegraphics[width=0.8\columnwidth]  {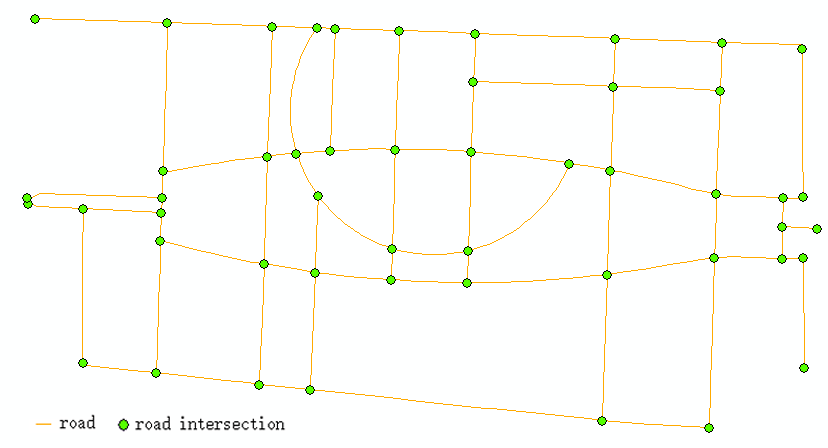}} 
	\caption{\label{1} Road network map.} 
\end{figure}
\par As shown in Figure 5, the road network can often be represented by an undirected graph. The nodes in this figure represent the end or intersections of the road in the road network. The line segment connected by two nodes includes attributes such as length and width. These attributes can describe the detailed information of each path in the map and the topology information between the paths clearly. Under epidemic control, students strictly abide by school regulations and start from their location to their destination within the permitted time along fixed route, which has no closed loop. For the moving trajectory of students from dormitory building to other areas, the starting point is taken as the root node, the midway road node is taken as the child node, and the destination is taken as the leaf node. A multi-branch tree with the dormitory building as the root node will be formed, as shown in Figure 6.

\begin{figure}[!t] 
	\center{\includegraphics[width=0.9\columnwidth]  {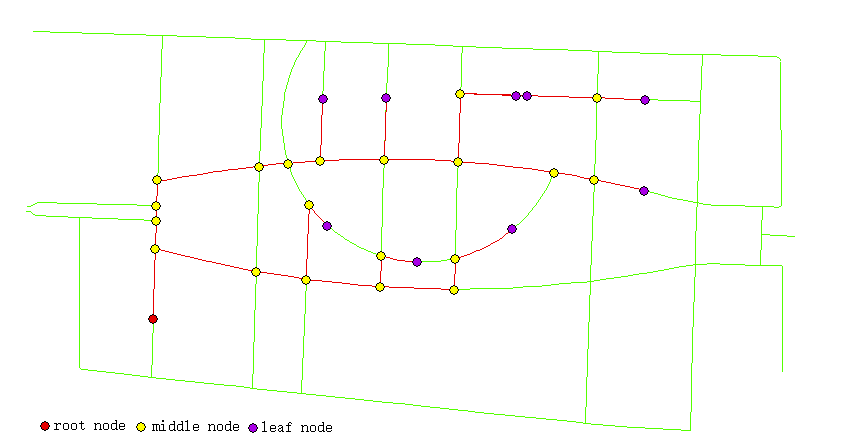}} 
	\caption{\label{1} Node classification  of road network.} 
\end{figure}
\par The multi-branch tree can be used to record the topology of each required road segment, and the attribute information of the road segment is described by the node attributes. The $Length$ attribute in Table 3 not only includes the length of the path itself, but also the distance from the root node to this one. For example, the distance from a node to its parent is $5.1m$, and the distance to its root node is $56.55m$. The $Weight$ attribute is a sequence, which contains the time when the crowd passes this road node and the corresponding number of the crowd, taking the weight attribute <300,600,0.05> as an example, it means that about $5\%$ of the total number of students will pass through this node between $300s$ and $600s$.

\begin{table}[!t]
	\caption{\textbf{Topological properties of the road sections}}
	\centering
	\begin{tabular}{m{2cm}<{\raggedright} p{5.5cm}}
		\toprule
		Attributes &  Description \\
		\midrule
		ParentsNode &  The parent node of this road node. \\
		ChildrenNode &All children nodes of this road node. \\
	    Length(m)&	The length of the path is represented by this node to its parent node.\\
		Width(m)&	The width of the path from this node to its parent node. \\
		Weight&	The total number of people passing by this node (percentage of total people).\\
		\bottomrule
	\end{tabular}
\end{table}
  The time interval of passing through the node by the crowd is inversely proportional to the width of the road. According to the weight attribute, the crowd density of any road node can be estimated at one certain moment. The time when the crowd pass through the road node can be expressed by the following formula:\\
\begin{equation}
\begin{array}{l}
t_{i(start)} = {l_{i,tatal}}/{{v_{max}}}\\
t_{i(end)} = {l_{i,tatal}}/{{v_{min}}}\\
\Delta t'_{i} = t_{i(end)} - t_{i(start)}
\end{array}
\end{equation}
where $t_{i(start)}$ indicates the starting time for the crowd to reach the $i$ road node, which is proportional to the distance from the node to the root;  $t_{i(end)}$ indicates the last time for the crowd to reach the $i$ road node; $l_{i,tatal}$ is the total distance from the starting point to node $i$; $v_{max}$ and $v_{min}$ respectively represent the maximum and minimum speed of the crowd; $\Delta t'_{i}$ indicates the difference between the earliest and latest time when the flow of people passes the path node.

\subsection{Crowd simulation and control}
\subsubsection{Crowd simulation}
Students depart from the dormitory, go to the target functional area in turn, and return to the dormitory after completing the visit. Under quarantine conditions, infected and suspected infected persons need to go to the school hospital for testing and isolation. Each agent is only allowed to move along a given road to the  destination at the permitted time, and the rest of the time is not allowed to walk around. Individuals may be infected inside dormitories, classes, restaurants, roads and so on.
\subsubsection{ Crowd control}
According to the characteristics of virus transmission, three control measures have been formulated from the perspectives of reducing population density and improving self-protection, such as batch travel (reduce aggregation by reducing the number of travelers at the same time), staggered travel (reduce road congestion by controlling travel time), and isolation prevention (keep suspected patients away from the crowd). Based on the  experimental simulation results, we analyze the impact of different measures on the results of the virus transmission.

\section{Experiment}
The experiment is based on $CPU$ $3.60 GHz$, $8GB$ memory, and Windows 10 operating system environment. We implement the simulation in C++ based on $PEDSIM$  platform \cite{gloor2016pedsim}. The system allows users to set parameters by themselves to calculate the infection curve, such as the initial total population, the initial number of patients and the number of days of virus transition for patients from different states, etc.

\par Since there is no large-scale outbreak of novel coronavirus in universities around the world in public reports, the real infection case can not be obtained. Therefore, we use the reported data of Diamond Princess in Japan as the reference to conduct our simulation. The reason why we use this case is as following: First, both Zhengzhou University and Diamond Princess Cruise are relatively closed environment. According to media reports, the passengers and crews on the cruise are not allowed to leave freely. Zhengzhou University also takes strict closed-end management to ensure the health of teachers and students at the beginning of epidemic. Both of these two actual situations are very similar. Second, they have similar categories of population and space occupancy. The population consists of staff and students at Zhengzhou University and the Diamond Princess Cruise mainly has crews and passengers. The movement of crowds on them is primarily on foot, and the speed of pedestrians is also consistent. Moreover, the space occupancy between Zhengzhou University and the Diamond Princess Cruise are also very similar. The cruise is 290 meters long and 37.5 meters wide, with 18 floors and a total area of about 200 000 square meters. During the outbreak of COVID-19, there were 3,711 members on the Diamond Princess. And the main campus of Zhengzhou University covers 2,844,00 square meters and holds about 40,000 people. Through the careful verification, the space occupancy rate of these two cases are very close. Third, human daily activities are similar. Students and staff have classes on campus and their daily route is from the dormitory to the teaching building or other fixed ones. The passenger on the cruise will have a walk on the board and some crews need to deliver meals to passengers’ dormitories. The chances of the virus spreading among these activities are similar. Although the population size is not exactly the same between the university and the cruise, it does not affect the trend of virus transmission.

\par We scale the number of teachers and students in Zhengzhou University proportionally. Assuming that there is already an infected patient on the campus on the first day, according to the infection situation of the groups in different periods, the approximate infection number and infection rates of all teachers and students on the campus are inferred. In the simulation, each agent is only allowed to move along the established road to a specific place at the specified time, and the rest of the time is not allowed to move around at will. Individuals may be infected in dormitories, classrooms, restaurants, and roads. The number of infections at the end of each day is obtained through the infection model, and corresponding control measures are implemented according to the results to reduce the infection rate. In order to prevent accidental errors in the experiment, our results are obtained through averaging multiple rounds of calculation.

\begin{figure}[!t] 
	
	\center{\includegraphics[width=1.0\linewidth]  {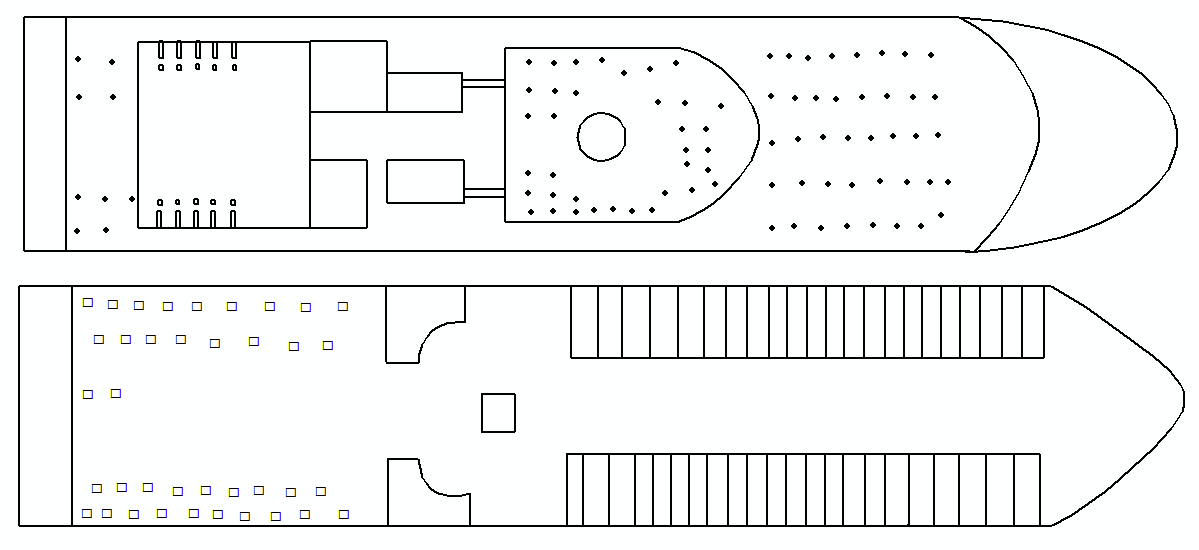}} 
	\caption{\label{1}  Sample scene of the second upper deck and  the sixth sun deck  of Diamond Princess.   } 
\end{figure}  

\subsection{Model validation}

Since COVID-19 did not actually happen in Zhengzhou University, in order to verify our proposed model, the similar case of Diamond Princess Cruise is involved in our experiment. Although there are some differences between cruise and university campus, they are essentially closed environment and have plenty of similarities as mentioned above. The first case of coronavirus on the cruise  was on Feb. 1st. A few days later, the cruise was ordered by the Japanese government to quarantine, and no one was allowed to leave the ship. The Diamond Princess is a special and typical infectious disease sample in the global COVID-19. It is also known as a highly infectious experimental model of COVID-19.

The scene modeling of the Diamond Princess deck is shown in Figure 7. In the experiment, passengers in public areas such as decks and restaurants can walk freely, and they are allowed to change floors to enter other spaces within a specified time. Most of the time, passengers can go to the designated room for leisure and entertainment.  At this time, the overall protection rate ($\beta$) is 0. We use the reported infection data of the Princess Diamond as the ground truth to validate our method, which is shown in the red curve in Figure 8. Our simulation result is shown in green curve. Orazio et al. \cite{d2020probabilistic} also used an agent-based model to simulate the situation of the Diamond Princess, whose result is shown in the blue curve. Through the comparison, it shows that our simulation result is reasonable and accurate.

\begin{figure}[!t] 
	\center{\includegraphics[width=0.85\columnwidth]  {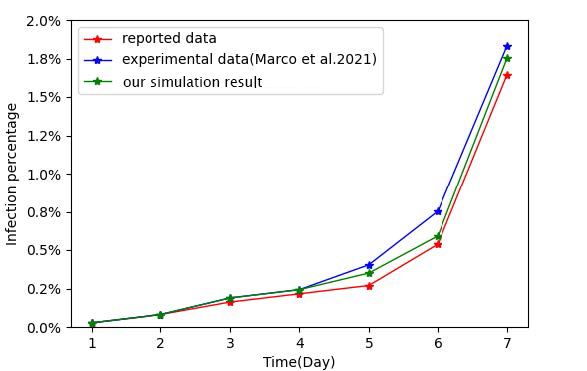}} 
	\caption{\label{1} The comparison between our result and others. } 
\end{figure}   

\subsection{Population size}
When the total number of people changes, the probability of contact between individuals, movement trends and other factors will change, which will have different impacts on the spread and infection of the virus. The population on campus is scaled to different proportions. The initial number of people is set to 840, 1260, 1680, and 2520 respectively, and the infection situation of the population within 15 days is calculated through the model.
\begin{figure}[!t] 
	\center{\includegraphics[width=0.85\columnwidth]  {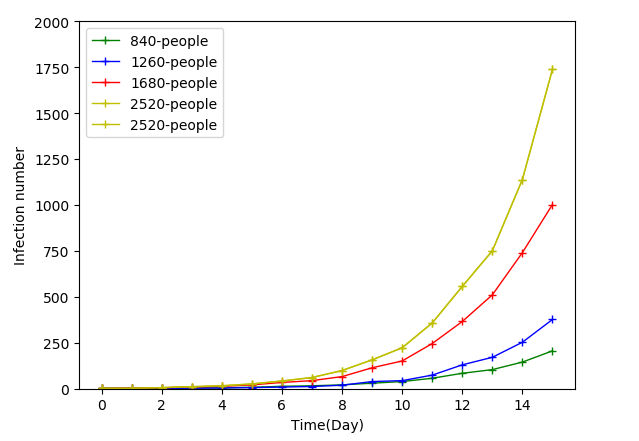}} 
	\caption{\label{1} Influence of different population sizes on virus transmission.} 
\end{figure}
\par The results in Figure 9 show that the larger the total number of people on campus, the more people will be infected eventually. If there are no restrictions, the number of infected people will increase significantly over time. However, this trend is not exactly proportional. When the density of campus population is too small, the infection rate will rise relatively slowly; when the population density is large, the infection rate will also increase significantly as the total population increases. 
\par Figure 10 shows the proportion of the number of people in each state per day when the total number of simulated amount is 1680. The B-lurker curve shows the change of the number of patients in the incubation period; the B-infected curve shows the change of the number of patients with confirmed infections; The w/o control curve represents the change of the total number of people diagnosed and incubated each day. It can be seen that the curve grows slowly at the beginning, and then breaks out. Almost all the people will be infected on the 21st day. Based on this, after students return to school, it is necessary to take corresponding protective control measures to reduce the impact of the virus.
\begin{figure}[!t] 
	\center{\includegraphics[width=0.85\columnwidth]  {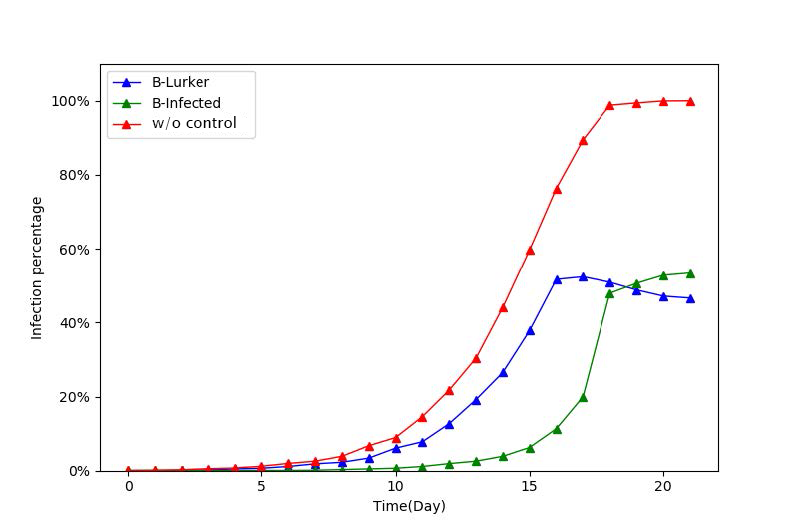}} 
	\caption{\label{1}  The infection rate curves in 21 days without control.} 
\end{figure}

\subsection{Control measures}
\subsubsection{Batch travel measures}
From the perspective of time and space, we formulate intervention measures to reduce the risk of transmission. The first control is to make students travel in batch. Since teachers and students have basically the same activity areas on campus, there will often be a large number of people gathering in some areas. Therefore, we should reduce the occurrence of such phenomena and avoid large crowd gathering. Classes are commonly the main component of the daily life for the college students, so first of all, according to the individual attributes, the class time is re-planned for the students  with course tasks. $50\%$ of the teachers and students with course tasks are arranged to have classes in the morning, and the other half of the students are arranged in the afternoon. In class time, individuals are only allowed to stay in the dormitory when they have no class tasks, and all individuals are not allowed to move around the campus at will. Planning student travel in batches in this way can greatly reduce the crowd density in most cases. For example, by reducing the number of people and expanding the physical space between students during class, it can reduce the crowding degree of students.

\begin{figure}[!t] 
	\center{\includegraphics[width=0.85\columnwidth]  {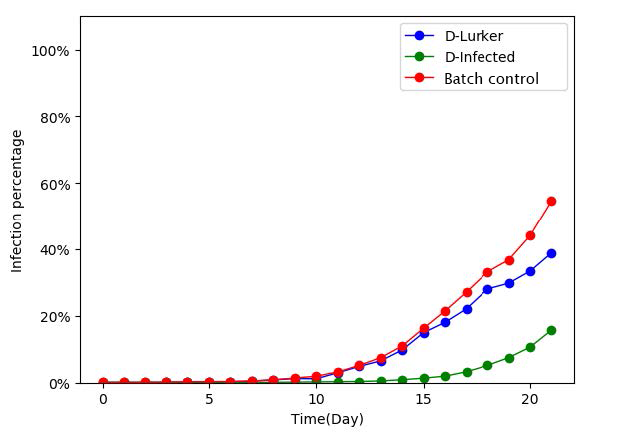}} 
	\caption{ \label{1}  The infection rate curves in 21 days using the control of batch travel.} 
\end{figure}

\par Figure 11 shows the comparison of the infection results of the population before and after the implementation of batch control. The D-lurker curve represents the change curve of the number of lurking people per day after the implementation of batch control; The D-infected curve represents the number of people diagnosed each day after the implementation of batch control; Batch control measures represent the total number of infections per day after the implementation of batch control. It is found that the upward trend of the curve is relatively flat, and the final total number of infections is reduced by about half compared with that before control. The effect of post-implementation measures is in line with expectations, and the spread of the virus can be suppressed to some extent.
\subsubsection{Staggered travel measures}
In previous section, we divide the crowd into two groups by replanning the class time, which avoided crowded contact to a certain extent. However, the university group is huge, under the same batch, students in the same dormitory building will still experience congestion on the road. In addition to classes, the restaurant is also a high-density place of crowded gathering. At the Mealtime, the students will flock to the restaurant. The crowd is very dense on the way to the restaurant and after arriving in the restaurant.
\par
According to the initial location and the destination  of each student, we calculate the specific travel time of the students in different dormitories. For example, when students move from the dormitory area to other areas, the dormitory building determines the travel time according to the location between itself and other buildings and also the path selected by the students. By controlling the travel time in different buildings, the road utilization at a certain moment can be reduced. Through our algorithm, we can easily calculate the reasonable departure time of the groups in each building area, reducing the mutual contact among the groups. 
\par Students are distributed in various dormitory buildings, and teaching areas are also distributed in different areas of the campus. Campus under epidemic control, multiple multitrees are constructed according to the route of all students according to their starting point. If two or more multitrees contain the same road node and the time overlaps, it means that the two paths may collide with pedestrian flows. It needs to stagger a little time, and adjust the departure time offset $\Delta t$ to reduce conflicts. If multiple groups pass the same node on different path trees, and there is no overlap in the elapsed time, it is not considered crowds will collide. 
\par If the road node $i$ appears at the same time in different paths and the passing time overlaps, we calculate the road congestion with the following formula: 
\begin{equation}
C = {\sum_{i = 0}^{n}{\sum_{j = 0}^{n_{1}}\frac{\omega_{ij}}{\Delta t'_{ij}}}}
\end{equation}
where $n$ represents the total number of road nodes in the map; $n_{1}$ indicates the total number of departure places for pedestrians; $\omega_{ij}$ indicates the weight of the flow of people from the $j$th starting place passing through the $i$th road node; $\Delta t_{ij}^{'}$ represents the time interval for the flow of people from the $j$th departure place to pass through the  $i$th road node. The offset time of each departure place is adjusted to the minimum value, as in the following formula:
\begin{equation}
min\left( {\sum\limits_{i}^{n}{\Delta  t_{ij}}} \middle| 0 < j < n_{1} \right)
\end{equation}
In our experiment, students are evenly distributed in 5 dormitory buildings. Students start from the dormitory building to other functional areas of the campus, and they are required to meet the constraints, that is, ${\mathit{\max}\left( {t_{i}} \right)} \leq 20min$. More ideal results can be obtained through brute force traversal and integer planning. The starting time for classes in different locations are shown in Table 4. 

\begin{table}[!t]
	\caption{\textbf{Travel time for class}}
	\centering
	\begin{tabular}{p{4cm}<{\centering} p{4cm}<{\centering} }
		\toprule
		Building & Start Time \\
		\midrule
		Dormitory Building 1&0s \\
		Dormitory Building 2&	600s\\
		Dormitory Building 3&	1080s\\
		Dormitory Building 4&	360s\\
		Dormitory Building 5&	120s\\
		\bottomrule
	\end{tabular}
\end{table}

\par After the course, students will move to their dormitories. According to the distribution of students, the class time is divided into three groups: teaching building, library, others (experimental building, administrative building). The result of the preparation time to get out of class is shown in Table 5.

\begin{table}[!t]
	\caption{\textbf{Travel time after class}}
	\centering
	\begin{tabular}{p{6cm}<{\centering}  p{1.5cm}<{\centering}}
		\toprule
		Building& Start Time \\
		\midrule
		Library&0s \\
		Teaching building&900s \\
		Others (laboratory building, administrative building)&360s \\
		\bottomrule
	\end{tabular}
\end{table}
\par The heat maps in Figure 12 show the crowd density under different conditions on the path.  We take Figure 12(a) as an example. Figure 12(b) means that there are three groups of people from different locations to the same direction at the same time. Figure 12(c) represents the density of people on each path that is not controlled. It can be seen that when pedestrians on multiple paths move toward the same destination at the same time, there will be a high density. Figure 12(d) shows that by adjusting the travel time, it enables students to implement non-intersection travel plans, which can effectively reduce the crowd density on the road. 

\begin{figure}[!t]
	\centering
	\subfigure[A sample of map]{
		\begin{minipage}[t]{0.45\columnwidth}
			\centering
			\includegraphics[width=1.51in]{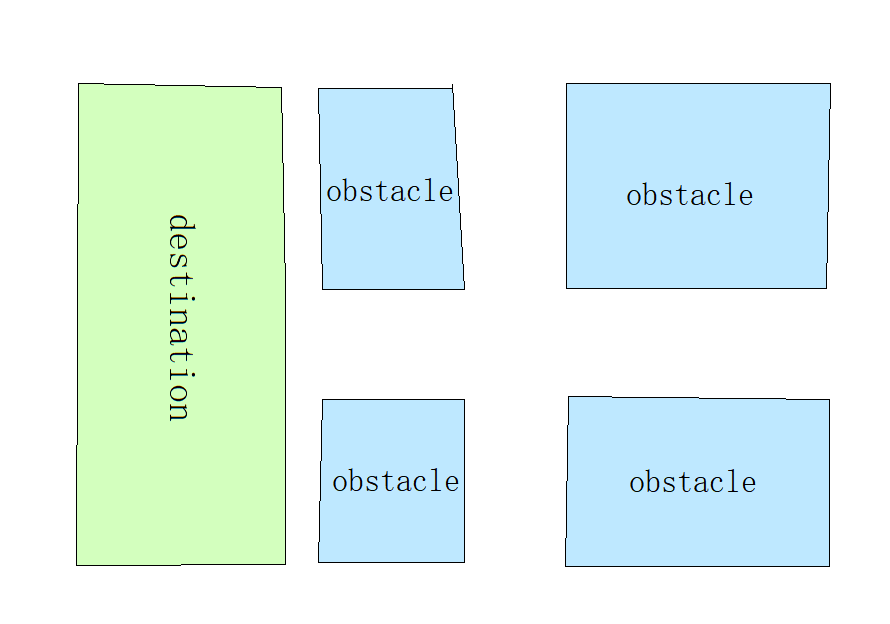}
		\end{minipage}%
	}%
	\subfigure[Path heat map after staggered traveling]{
		\begin{minipage}[t]{0.45\columnwidth}
			\centering
			\includegraphics[width=1in]{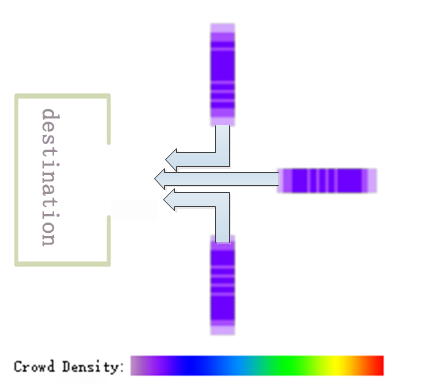}
		\end{minipage}
	}%
	
	\subfigure[Before staggered traveling]{
		\begin{minipage}[t]{0.45\columnwidth}
			\centering
			\includegraphics[width=1in]{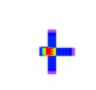}
		\end{minipage}%
	}%
	\subfigure[After staggered traveling]{
		\begin{minipage}[t]{0.45\columnwidth}
			\centering
			\includegraphics[width=1in]{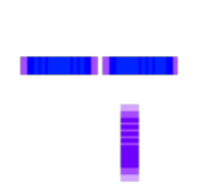}
		\end{minipage}
	}%
	\centering
	\caption{ Heat map of path density before and after staggered traveling.}
\end{figure}

\begin{figure}[!t] 
	\center{\includegraphics[width=0.85\columnwidth]  {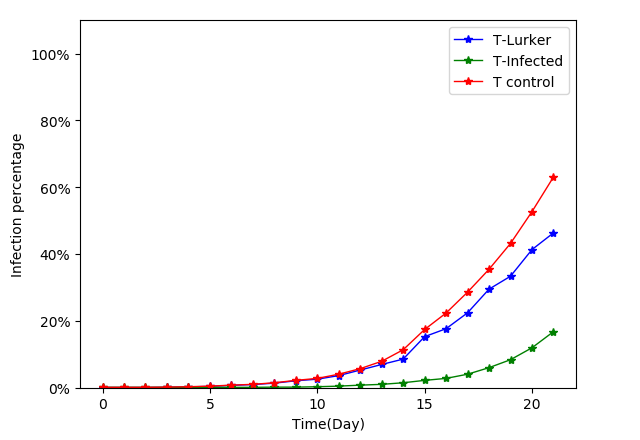}} 
	\caption{\label{1} The infection rate curves in 21 days.} 
\end{figure}

\par As shown in Figure 13, T-control represents the change curve of the total number of persons in the incubation period and infected persons every day after the implementation of staggered travel control measures. It can be seen that the increasing trend is slow compared with the other two curves. By implementing control measures to reduce the population density, the probability of virus transmission can be further reduced.

\subsubsection{Isolation control measures}

The traditional method of interviewing infected people to track contacts is not so effective. We use contact tracking method to locate infected people as \cite{shamil2021agent}. On campus,
smartphones have almost become the necessary supplies for college students. Teachers and students on campus should install contact tracking applications on these electronic devices. It is a good way to track (and then isolate) individuals who have been in close contact with infected patients before symptoms appear. Because of the high infectiousness, once a confirmed patient is found, the person should be isolated immediately \cite{chen2020time}. If one confirmed patient is found on the campus, the student will be immediately sent to the school hospital for closed isolation. By tracing the trajectory of this student, those who have been in close contact with him/her are regarded as suspected cases and will be sent to the school hospital for isolation and observation as in \cite{chowell2004basic}.

\par However, since some virus carriers may have no obvious symptoms of infection after the incubation period, who are asymptomatic patients. Such patients need to rely on medical methods to determine their physical health, which means the difficulty of epidemic investigation and control will be further increased. Three sets of experiments are conducted including the situation where there are asymptomatic patients. 
\begin{figure}[!t] 
	\center{\includegraphics[width=0.85\columnwidth]  {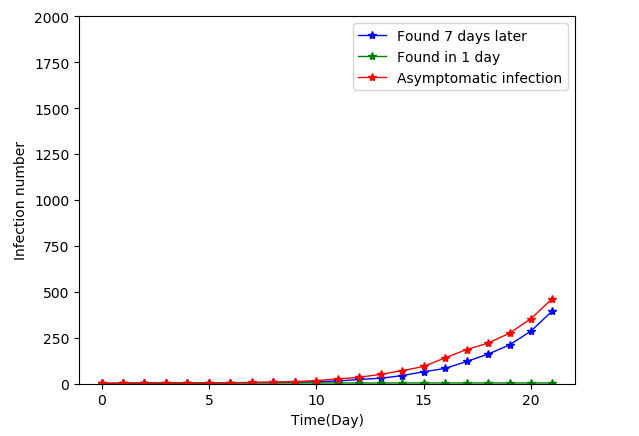}} 
	\caption{\label{1} The infection rate curves in 21 days under different disease states.} 
\end{figure}

\par The results are shown in Figure 14, the blue curve is the trend of the infection rate  with the virus on the first day and being isolated and treated after 7 days of infection; The green curve shows the trend of the infection rate of the population when one person is infected on the first day and isolated in time; The red curve represents the daily infection trend when  one person is infected with the virus on the first day and all patients have a probability of 0.1 as asymptomatic infections. The appearance of asymptomatic patients will make the condition more complicate. Only by detecting all virus carriers can it be possible to completely control the spread of the virus, otherwise the virus will still spread on a large scale.

\begin{figure}[!t] 
	\center{\includegraphics[width=0.85\columnwidth]  {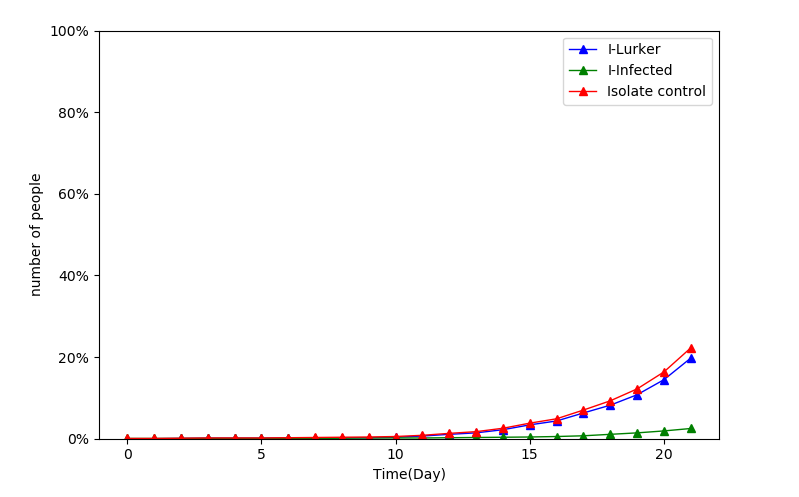}} 
	\caption{\label{1} The infection rate curves in 21 days after the implementation of isolation control measures.} 
\end{figure}

\par As shown in Figure 15, the three curves of I-lurker, I-infected, and isolate control respectively represent the changing trends of the number of suspected cases, the number of confirmed persons, and the total number of infections after the implementation of mandatory isolation. The number of infected people each day after the quarantine measures is less than that before the implementation of the quarantine measures, and the total number of people infected on the 21st day is only $24\%$. Compared with the control measures in previous two sections, it is proved that the implementation of isolation measures can more effectively reduce the probability of people infected with the virus. Isolation control measures can effectively reduce the spread rate of the virus, but it does not completely contain the virus. The reason may be that individuals are infected by the virus in the moving on the road, but the infected people are not located accurately.

\begin{figure}[!t] 
	\center{\includegraphics[width=0.85\columnwidth]  {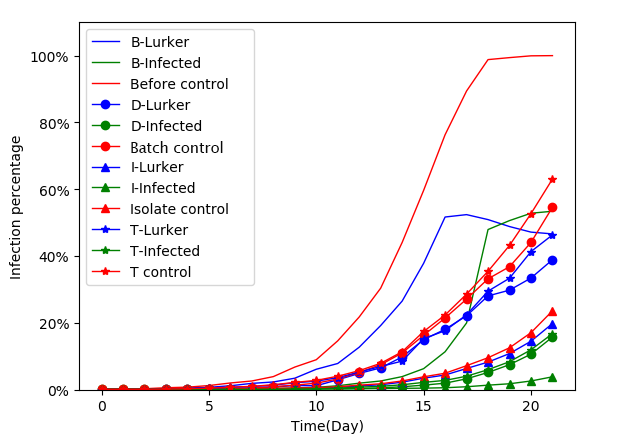}} 
	\caption{\label{1} The comparison result without control, batching, staggered travel, and isolation control measures.} 
\end{figure}

\par  As shown in Figure 16, experimental results show that in a high-density and closed area such as colleges and universities, even if only few agents are infected with the virus, the infection probability of the crowd will still be higher without any control measures. After taking compulsory isolation and treatment measures, the spread of the virus will be obviously slowed down.

\subsection{ Proportion of self-protection population}

Since the strong transmission of the novel coronavirus, people begin to realize the importance of self-protection. Due to the differences of individual immune system, the infection probability of each person is not exactly the same. In addition to the protection measures implemented by universities on teachers and students, individuals also need to strengthen self-protection. For example, wearing a mask and keeping a distance from others consciously. Once the group adopts protective measures, the spread of the virus will be reduced, and the group infection rate will also decrease dramatically. 

\begin{figure}[!t] 
	\center{\includegraphics[width=0.85\columnwidth]  {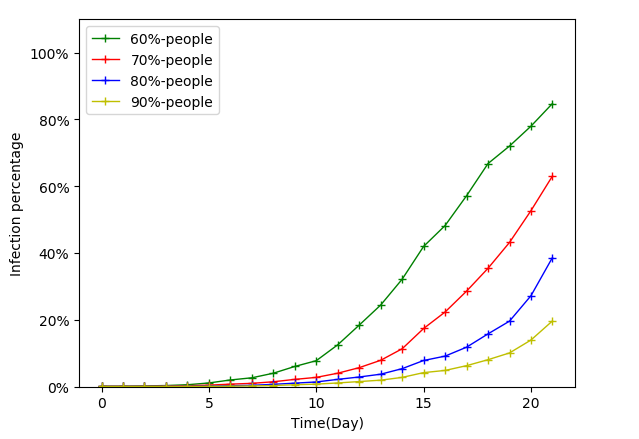}} 
	\caption{\label{1} The infection curve with different self-protection rates.} 
\end{figure}

\par Figure 17 shows the impact of different self-protection rates in groups in universities on the virus infection rate within 21 days. The green, red, blue, and yellow curves represent the daily viral infection rate of the population when the population protection rate is $60\%$, $70\%$, $ 80\%$, and $ 90\%$. On the 21st day, when $60\%$ of the people in colleges and universities take self-protection measures, the green curve shows that the virus infection rate is $82\%$. The result is much higher than that under the same conditions where the number of people taking protective measures reaches $70\%$ and above. The more people who take protective measures, the lower the infection rate is. Therefore, universities should actively call on all teachers and students to take self-protections for the prevention of virus spreading.

\section{Conclusion}
In response to the outbreak of COVID-19, the governments in different countries have implemented a series of measures to prevent a large number of people from gathering and aggravating the epidemic. Taking the problem of virus transmission after students return to school as an example, we propose a Campus Virus Infection and Control Simulation (CVICS) model that is oriented to a closed environment, frequent crowd contact, and strong mobility. By constructing an agent-based simulation, and taking into account the differences among individuals, we introduce the mean field theory to micro-simulate the infection situation of each individual every day. The experimental results show that our model can calculate the daily population infection trend and individual infection status. Through batch travel, staggered travel, isolation and other effective and efficient control measures, it can reduce the probability of population infection and curb the rapid spread of the virus to relatively low level. During the epidemic, a series of tough measures should be taken to reduce crowd gathering. Once suspected patients are found, compulsory measures should be taken to isolate them under medical observation. When the isolation is strong, the source of infection is blocked, and the spread of novel coronavirus will be better controlled. For individuals, we should try our best to avoid crowd gathering, reduce the frequent crowd contact, and enhance self-protection. 
\par In the future, we will consider more environmental factors such as bicycle, car and others in the simulation. And more convincing medical models will be integrated in the infection model to enhance the authority. We not only should consider these existing control measures, but also resource allocation, such as adding disinfection and vaccine measures, to further improve the accuracy of virus infection model. The improved model should be able to simulate more complex crowd flow in large-scale areas, and demonstrate real infection data as much as possible. In view of the current epidemic prevention and control in large cities in China, such as Nanjing and Zhengzhou, the local governments classified different regions into closed area, sealed area, prevention and controlled area. The people in the closed area are not allowed to go out of their own house and the community will provide them with the necessary living supports. The residents in the sealed area are strictly prohibited gathering and cannot go out of their community freely. And the inhabitants in prevention and control area, they are required to stay at home most of the time  and cannot leave unless emergency conditions. People in various regions must accept different ways of nucleic acid tests, such as one by one or ten person as one group, which can help to locate the infected person quicker and more accurately from tens of millions of people in large cities. Above measures have been proved very efficient for large cities, however, they are not very suitable and necessary for the campus. In the future, we will try to simulate the above measures for large cities. This will make the simulation of the epidemic prevention and control more flexible to areas of different sizes.


%



%
\bibliographystyle{IEEEtran}
\bibliography{IEEEabrv,paper}

\begin{thebibliography}{10}
\providecommand{\url}[1]{#1}
\csname url@samestyle\endcsname
\providecommand{\newblock}{\relax}
\providecommand{\bibinfo}[2]{#2}
\providecommand{\BIBentrySTDinterwordspacing}{\spaceskip=0pt\relax}
\providecommand{\BIBentryALTinterwordstretchfactor}{4}
\providecommand{\BIBentryALTinterwordspacing}{\spaceskip=\fontdimen2\font plus
\BIBentryALTinterwordstretchfactor\fontdimen3\font minus
  \fontdimen4\font\relax}
\providecommand{\BIBforeignlanguage}[2]{{%
\expandafter\ifx\csname l@#1\endcsname\relax
\typeout{** WARNING: IEEEtran.bst: No hyphenation pattern has been}%
\typeout{** loaded for the language `#1'. Using the pattern for}%
\typeout{** the default language instead.}%
\else
\language=\csname l@#1\endcsname
\fi
#2}}
\providecommand{\BIBdecl}{\relax}
\BIBdecl

\bibitem{world2020global}
W.~H. Organization \emph{et~al.}, ``Global research on coronavirus disease
  (covid-19),'' \emph{World Health Organzation [internet]}, 2020.

\bibitem{milne2008small}
G.~J. Milne, J.~K. Kelso, H.~A. Kelly, S.~T. Huband, and J.~McVernon, ``A small
  community model for the transmission of infectious diseases: comparison of
  school closure as an intervention in individual-based models of an influenza
  pandemic,'' \emph{PloS one}, vol.~3, no.~12, p. e4005, 2008.

\bibitem{li1995global}
M.~Y. Li and J.~S. Muldowney, ``Global stability for the seir model in
  epidemiology,'' \emph{Mathematical biosciences}, vol. 125, no.~2, pp.
  155--164, 1995.

\bibitem{ridenhour2011controlling}
B.~J. Ridenhour, A.~Braun, T.~Teyrasse, and D.~Goldsman, ``Controlling the
  spread of disease in schools,'' \emph{PloS one}, vol.~6, no.~12, p. e29640,
  2011.

\bibitem{tabari2020nations}
P.~Tabari, M.~Amini, M.~Moghadami, and M.~Moosavi, ``Nations’ responses and
  control measures in confrontation with the novel coronavirus disease
  (covid-19) outbreak: A rapid review,'' 2020.

\bibitem{fischer2009gpu}
L.~G. Fischer, R.~Silveira, and L.~Nedel, ``Gpu accelerated path-planning for
  multi-agents in virtual environments,'' in \emph{2009 VIII Brazilian
  Symposium on Games and Digital Entertainment}.\hskip 1em plus 0.5em minus
  0.4em\relax IEEE, 2009, pp. 101--110.

\bibitem{1995Social}
D.~Helbing and P.~Molnar, ``Social force model for pedestrian dynamics,''
  \emph{Phys.rev.e}, vol.~51, no.~5, p. 4282, 1995.

\bibitem{zanlungo2011social}
F.~Zanlungo, T.~Ikeda, and T.~Kanda, ``Social force model with explicit
  collision prediction,'' \emph{EPL (Europhysics Letters)}, vol.~93, no.~6, p.
  68005, 2011.

\bibitem{towers2009pandemic}
S.~Towers and Z.~Feng, ``Pandemic h1n1 influenza: predicting the course of a
  pandemic and assessing the efficacy of the planned vaccination programme in
  the united states,'' \emph{Eurosurveillance}, vol.~14, no.~41, p. 19358,
  2009.

\bibitem{liang2020mathematical}
K.~Liang, ``Mathematical model of infection kinetics and its analysis for
  covid-19, sars and mers,'' \emph{Infection, Genetics and Evolution}, p.
  104306, 2020.

\bibitem{cunde2002si}
Y.~Cunde and H.~Baoan, ``A si epidemic model with two-stage structure,''
  \emph{Acta Mathematicae Applicatae Sinica}, no.~2, p.~02, 2002.

\bibitem{pang2007delayed}
G.~Pang and L.~Chen, ``A delayed sirs epidemic model with pulse vaccination,''
  \emph{Chaos, Solitons \& Fractals}, vol.~34, no.~5, pp. 1629--1635, 2007.

\bibitem{yu2018dynamic}
Y.~Yu, Y.~Shi, and W.~Yao, ``Dynamic model of tuberculosis considering
  multi-drug resistance and their applications,'' \emph{Infectious Disease
  Modelling}, vol.~3, pp. 362--372, 2018.

\bibitem{fisman2014early}
D.~Fisman, E.~Khoo, and A.~Tuite, ``Early epidemic dynamics of the west african
  2014 ebola outbreak: estimates derived with a simple two-parameter model,''
  \emph{PLoS currents}, vol.~6, 2014.

\bibitem{huerre2010simulating}
S.~Huerre, J.~Lee, M.~Lin, and C.~O'Sullivan, ``Simulating believable crowd and
  group behaviors,'' in \emph{ACM SIGGRAPH ASIA 2010 Courses}, 2010, pp. 1--92.

\bibitem{aschwanden2011empiric}
G.~D. Aschwanden, S.~Haegler, F.~Bosch{\'e}, L.~Van~Gool, and G.~Schmitt,
  ``Empiric design evaluation in urban planning,'' \emph{Automation in
  construction}, vol.~20, no.~3, pp. 299--310, 2011.

\bibitem{feng2016crowd}
T.~Feng, L.-F. Yu, S.-K. Yeung, K.~Yin, and K.~Zhou, ``Crowd-driven mid-scale
  layout design.'' \emph{ACM Trans. Graph.}, vol.~35, no.~4, pp. 132--1, 2016.

\bibitem{helbing1995social}
D.~Helbing and P.~Molnar, ``Social force model for pedestrian dynamics,''
  \emph{Physical review E}, vol.~51, no.~5, p. 4282, 1995.

\bibitem{huang2009microscopic}
X.-f. HUANG, K.-j. WANG, L.-y. GUO, and Q.~Shao, ``Microscopic simulation model
  study on pedestrian evacuation based on agent technology,'' \emph{Journal of
  System Simulation}, vol.~21, no.~15, pp. 4568--4571, 2009.

\bibitem{kleczkowski1999mean}
A.~Kleczkowski and B.~T. Grenfell, ``Mean-field-type equations for spread of
  epidemics: The ‘small world’model,'' \emph{Physica A: Statistical
  Mechanics and its Applications}, vol. 274, no. 1-2, pp. 355--360, 1999.

\bibitem{eubank2004modelling}
S.~Eubank, H.~Guclu, V.~A. Kumar, M.~V. Marathe, A.~Srinivasan, Z.~Toroczkai,
  and N.~Wang, ``Modelling disease outbreaks in realistic urban social
  networks,'' \emph{Nature}, vol. 429, no. 6988, pp. 180--184, 2004.

\bibitem{barrett2008episimdemics}
C.~L. Barrett, K.~R. Bisset, S.~G. Eubank, X.~Feng, and M.~V. Marathe,
  ``Episimdemics: an efficient algorithm for simulating the spread of
  infectious disease over large realistic social networks,'' in \emph{SC'08:
  Proceedings of the 2008 ACM/IEEE Conference on Supercomputing}.\hskip 1em
  plus 0.5em minus 0.4em\relax IEEE, 2008, pp. 1--12.

\bibitem{bissett2016integrated}
K.~Bissett, J.~Cadena, M.~Khan, C.~J. Kuhlman, B.~Lewis, and P.~A. Telionis,
  ``An integrated agent-based approach for modeling disease spread in large
  populations to support health informatics,'' in \emph{2016 IEEE-EMBS
  International Conference on Biomedical and Health Informatics (BHI)}.\hskip
  1em plus 0.5em minus 0.4em\relax IEEE, 2016, pp. 629--632.

\bibitem{yang2018mean}
Y.~Yang, R.~Luo, M.~Li, M.~Zhou, W.~Zhang, and J.~Wang, ``Mean field
  multi-agent reinforcement learning,'' \emph{arXiv preprint arXiv:1802.05438},
  2018.

\bibitem{2019Force}
Q.~Chao, X.~Jin, H.~W. Huang, S.~Foong, and S.~K. Yeung, ``Force-based
  heterogeneous traffic simulation for autonomous vehicle testing,'' in
  \emph{2019 International Conference on Robotics and Automation (ICRA)}, 2019.

\bibitem{0A}
Y.~Han, Q.~Chao, and X.~Jin, ``A simplified force model for mixed traffic
  simulation,'' \emph{Computer Animation and Virtual Worlds}.

\bibitem{deng2012fuzzy}
Y.~Deng, Y.~Chen, Y.~Zhang, and S.~Mahadevan, ``Fuzzy dijkstra algorithm for
  shortest path problem under uncertain environment,'' \emph{Applied Soft
  Computing}, vol.~12, no.~3, pp. 1231--1237, 2012.

\bibitem{li2020early}
Q.~Li, X.~Guan, P.~Wu, X.~Wang, L.~Zhou, Y.~Tong, R.~Ren, K.~S. Leung, E.~H.
  Lau, J.~Y. Wong \emph{et~al.}, ``Early transmission dynamics in wuhan, china,
  of novel coronavirus--infected pneumonia,'' \emph{New England Journal of
  Medicine}, 2020.

\bibitem{world2020management}
W.~H. Organization \emph{et~al.}, ``Management of ill travellers at points of
  entry--international airports, seaports and ground crossings--in the context
  of covid-19 outbreak: interim guidance, 16 february 2020,'' World Health
  Organization, Tech. Rep., 2020.

\bibitem{fang2020many}
Z.~Fang, Z.~Huang, X.~Li, J.~Zhang, W.~Lv, L.~Zhuang, X.~Xu, and N.~Huang,
  ``How many infections of covid-19 there will be in the" diamond
  princess"-predicted by a virus transmission model based on the simulation of
  crowd flow,'' \emph{arXiv preprint arXiv:2002.10616}, 2020.

\bibitem{liao2005probabilistic}
C.-M. Liao, C.-F. Chang, and H.-M. Liang, ``A probabilistic transmission
  dynamic model to assess indoor airborne infection risks,'' \emph{Risk
  Analysis: An International Journal}, vol.~25, no.~5, pp. 1097--1107, 2005.

\bibitem{gloor2016pedsim}
C.~Gloor, ``Pedsim: Pedestrian crowd simulation,'' \emph{URL http://pedsim.
  silmaril. org}, vol.~5, no.~1, 2016.

\bibitem{d2020probabilistic}
M.~D'Orazio, G.~Bernardini, and E.~Quagliarini, ``A probabilistic model to
  evaluate the effectiveness of main solutions to covid-19 spreading in
  university buildings according to proximity and time-based consolidated
  criteria,'' 2020.

\bibitem{shamil2021agent}
M.~S. Shamil, F.~Farheen, N.~Ibtehaz, I.~M. Khan, and M.~S. Rahman, ``An
  agent-based modeling of covid-19: validation, analysis, and
  recommendations,'' \emph{Cognitive Computation}, pp. 1--12, 2021.

\bibitem{chen2020time}
Y.~Chen, J.~Cheng, Y.~Jiang, and K.~Liu, ``A time delay dynamic system with
  external source for the local outbreak of 2019-ncov,'' \emph{Applicable
  Analysis}, pp. 1--12, 2020.

\bibitem{chowell2004basic}
G.~Chowell, N.~W. Hengartner, C.~Castillo-Chavez, P.~W. Fenimore, and J.~M.
  Hyman, ``The basic reproductive number of ebola and the effects of public
  health measures: the cases of congo and uganda,'' \emph{Journal of
  theoretical biology}, vol. 229, no.~1, pp. 119--126, 2004.

\end{thebibliography}



%

\newpage
\begin{IEEEbiography}[{\includegraphics[width=1in,height=1.25in,clip,keepaspectratio]{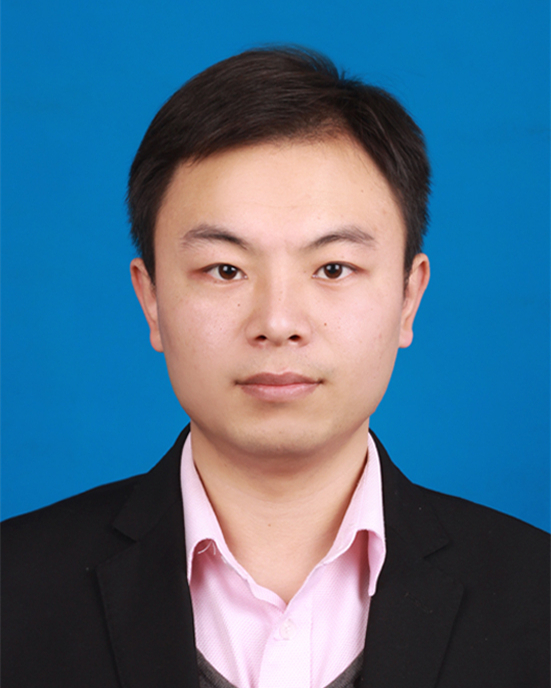}}]
{Pei Lv} received the Ph.D. degree from the State Key Laboratory of CAD\&CG, Zhejiang University Hangzhou, China, in 2013. He is an Associate Professor with the School
of Information Engineering, Zhengzhou University, Zhengzhou, China. His research interests include computer vision and computer graphics. He has authored more than 30 journal and conference papers in the above areas, including the IEEE T RANSACTIONS ON I MAGE P ROCESSING , the IEEE T RANSACTIONS ON C IRCUITS AND SYSTEMS FOR VIDEO TECHNOLOGY, CVPR, ACM MM, and IJCAI.
\end{IEEEbiography}

\begin{IEEEbiography}[{\includegraphics[width=1in,height=1.25in,clip,keepaspectratio]{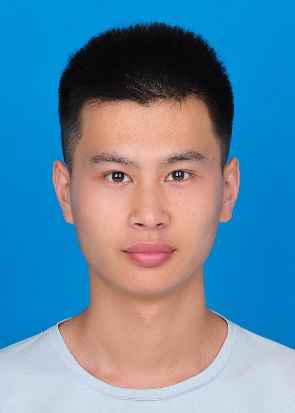}}]
{Quan Zhang} received the B.S. degree from the Computer Science and Technology Department, Putian University, China, in 2018. He is currently a master student in the School of Information Engineering of the Zhengzhou University. His research interests include computer graphics, crowd simulation.
\end{IEEEbiography}


\begin{IEEEbiography}[{\includegraphics[width=1in,height=1.25in,clip,keepaspectratio]{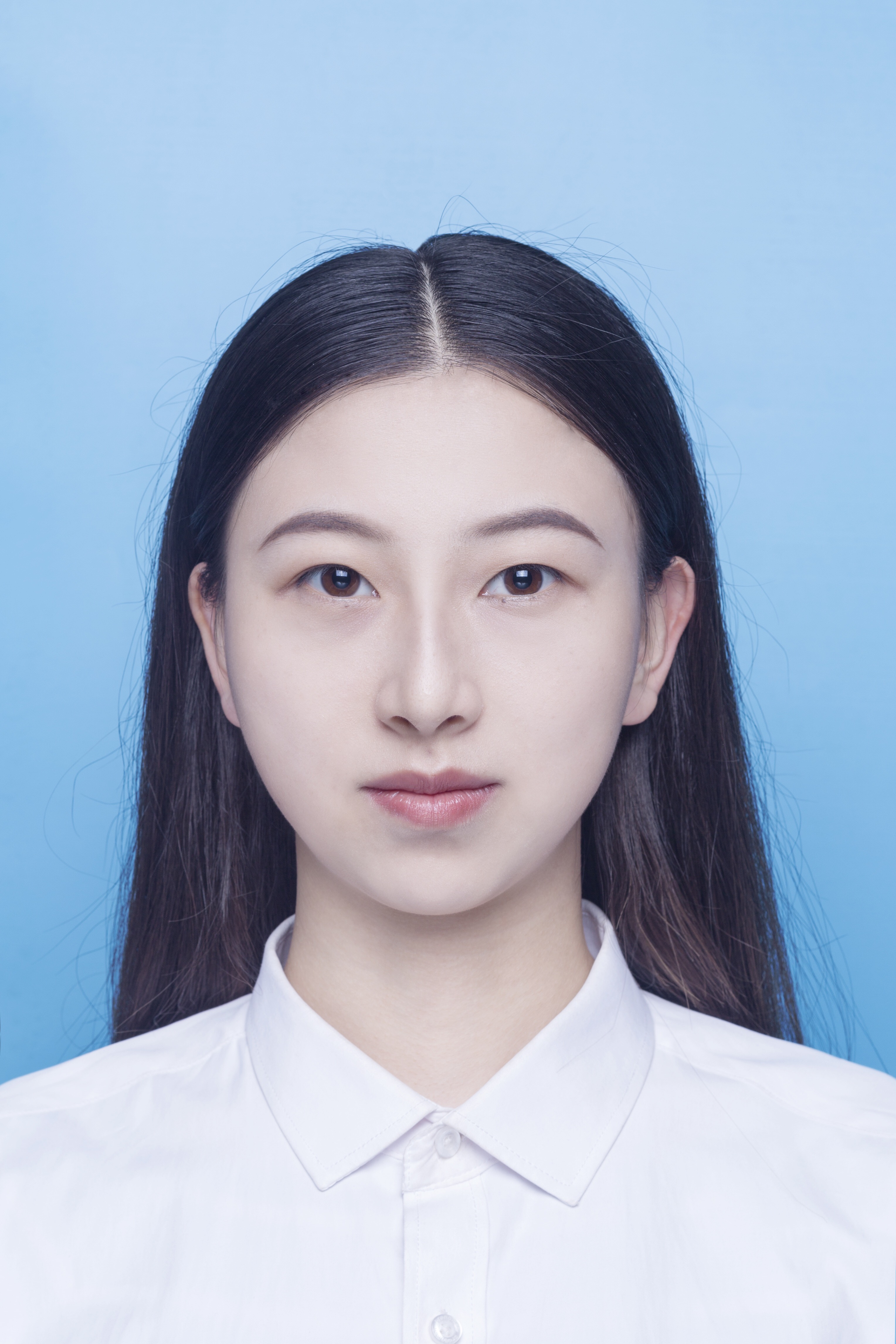}}]
{Boya Xu} received the B.S. degree from the Software Engineering Department, Zhengzhou University, China, in 2018. She is currently a master student in the School of Information Engineering of the Zhengzhou University. Her research interests include computer graphics, crowd simulation.
\end{IEEEbiography}

\begin{IEEEbiography}[{\includegraphics[width=1in,height=1.25in,clip,keepaspectratio]{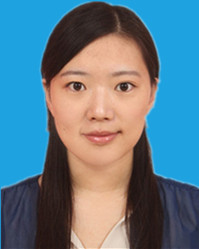}}]
{Ran Feng} received her B.Sc degree in Computer Science and Technology from Zhengzhou University. Zhengzhou, China, in 2012 and M.Sc degree in Computer Science from The University of Hong Kong, China, in 2015. She is currently a Ph.D. student in  the School of Information Engineering of Zhengzhou University. Her research interests include evolutionary computation and multiobjective optimization.
\end{IEEEbiography}

\begin{IEEEbiography}[{\includegraphics[width=1in,height=1.25in,clip,keepaspectratio]{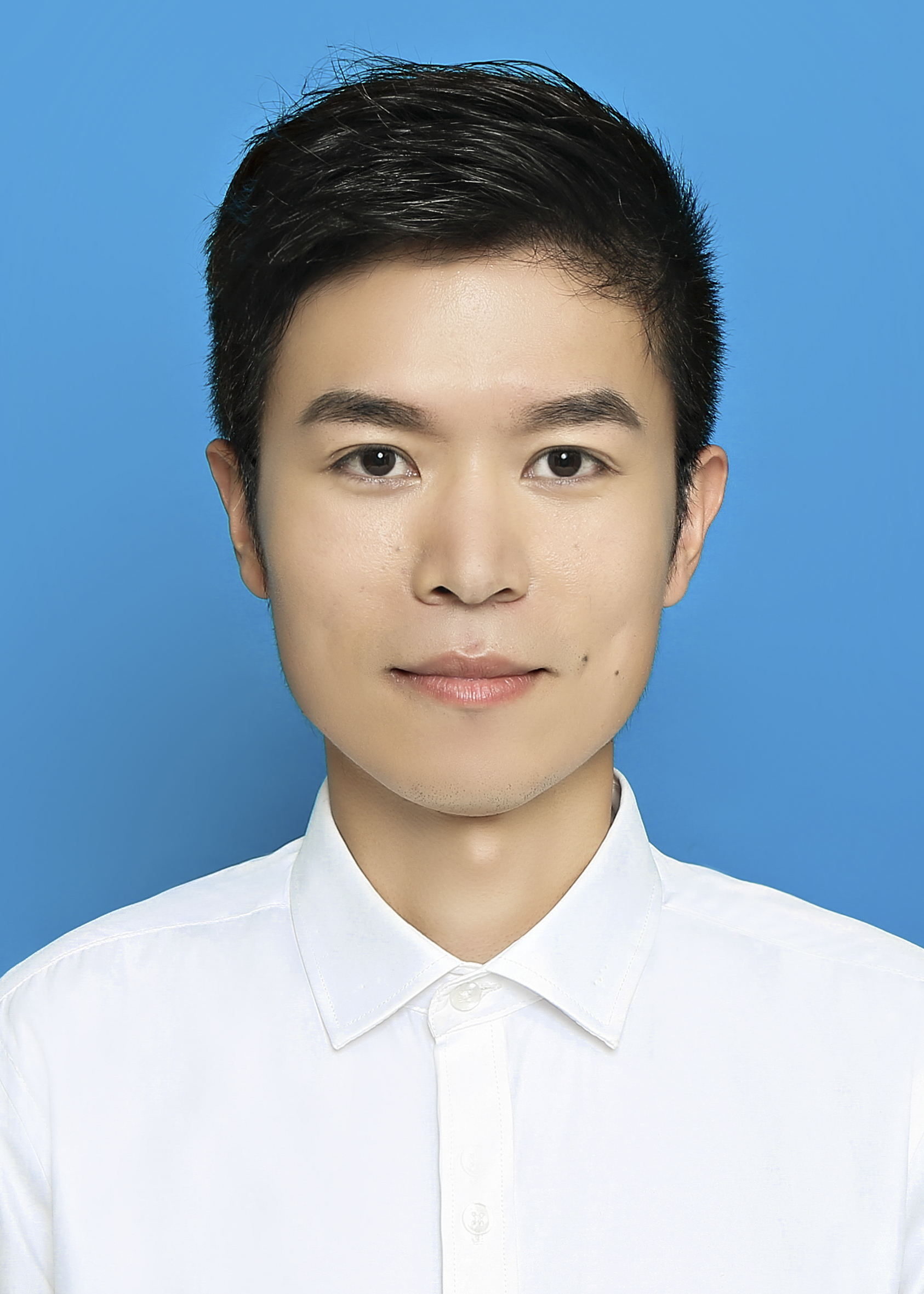}}]
{Chaochao Li} received his Ph.D. degree from the School of Information Engineering, Zhengzhou University, Zhengzhou, China. His current research interests include computer graphics and computer vision. He is currently an assistant research fellow with the School of Information Engineering, Zhengzhou University, Zhengzhou, China. He has authored over 6 journal and conference papers including the IEEE TRANSACTIONS ON AFFECTIVE COMPUTING, IEEE TRANSACTIONS ON INTELLIGENT TRANSPORTATION SYSTEMS, and IEEE TRANSACTIONS ON SYSTEMS, MAN, AND CYBERNETICS: SYSTEMS.
\end{IEEEbiography}

\begin{IEEEbiography}[{\includegraphics[width=1in,height=1.25in,clip,keepaspectratio]{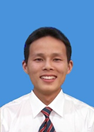}}]
{Junxiao Xue} is an associate professor in the School of Software of Zhengzhou University, China. His research interests include virtual reality and computer graphics. He received his Ph.D in 2009 from the School of Mathematical Sciences, Dalian University of Technology, China.
\end{IEEEbiography}

\begin{IEEEbiography}[{\includegraphics[width=1in,height=1.25in,clip,keepaspectratio]{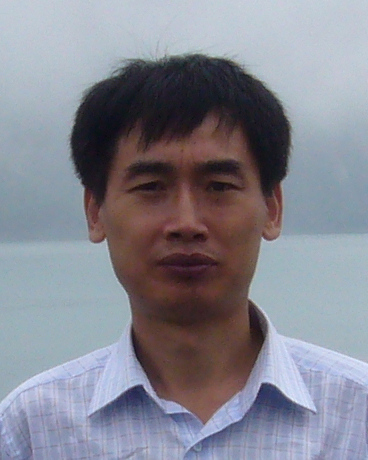}}]
{Bing Zhou} received the B.S. and M.S. degrees in computer science from Xian Jiao Tong University, Xian, China, in 1986 and 1989, respectively, and the Ph.D. degree in computer science from Beihang University, Beijing, China, in 2003. He is currently a Professor with the School of Information Engineering, Zhengzhou University, Zhengzhou, China. His research interests include video processing and understanding, surveillance, computer vision, and multimedia applications.
\end{IEEEbiography}

\begin{IEEEbiography}[{\includegraphics[width=1in,height=1.25in,clip,keepaspectratio]{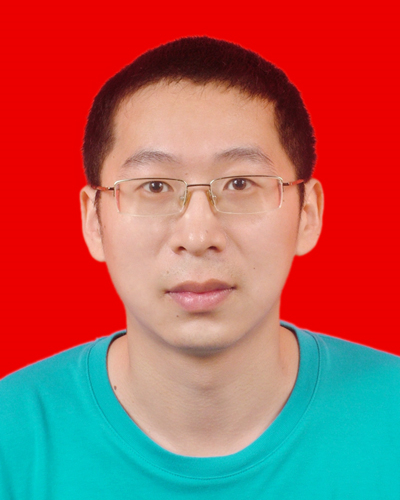}}]
{Mingliang Xu} received the Ph.D. degree in computer science and technology from the State Key Laboratory of CAD\&CG, Zhejiang University, Hangzhou, China, in 2012. He is a Full Professor with the School of Information Engineering, Zhengzhou University, Zhengzhou, China, where he is currently the Director of the Center for Interdisciplinary Information Science Research and the Vice General Secretary of ACM SIGAI China. His research interests include computer graphics, multimedia, and artificial intelligence. He has authored more than 60 journal and conference papers in the above areas, including the ACM Transactions on Graphics, the ACM Transactions on Intelligent Systems and Technology, the IEEE T RANSACTIONS ON P ATTERN A NALYSIS AND M ACHINE I NTELLIGENCE , the IEEE T RANSACTIONS ON I MAGE P ROCESSING , the IEEE T RANSACTIONS ON C YBERNETICS , the IEEE T RANSACTIONS ON C IRCUITS AND S YSTEMS FOR V IDEO T ECHNOLOGY , ACM SIGGRAPH (Asia), ACM MM, and ICCV.
	
\end{IEEEbiography}




\end{document}